\documentclass[12pt]{article}
\usepackage{amssymb,amsmath,cite}
\long\def\@makefntext#1{\parindent 0cm\noindent \hbox to
1em{\hss$^{\@thefnmark}$}#1}

\begin{document}
\begin{titlepage}
\vspace{.5in}
\begin{flushright}
\end{flushright}
\vspace{.5in}
\begin{center}
{\Large\bf
 Relativistic models of magnetars: Nonperturbative analytical approach  }\\
\vspace{.4in}
{Stoytcho~ S.~Yazadjiev$^{1,2}$\footnote{\it email: yazad@phys.uni-sofia.bg}\\
       {\footnotesize\it ${}^{1}$ \it Department of Theoretical Physics, Faculty of Physics,}
       {\footnotesize \it Sofia University, Sofia, 1164, Bulgaria }\\
            { \footnotesize  ${}^{2}$ \it Theoretical Astrophysics, Eberhard-Karls University of T\"ubingen, T\"ubingen 72076, Germany }}
\end{center}

\vspace{.5in}

\begin{abstract}
{In the present paper we focus  on building  simple nonperturbative
analytical relativistic models of magnetars. With this purpose in
mind we first develop a method for generating exact interior
solutions to the static and axisymmetric
Einstein-Maxwell-hydrodynamic equations with anisotropic perfect
fluid and with pure poloidal magnetic field. Then using an explicit
exact solution we present a simple magnetar model and calculate some
physically interesting quantities as the surface elipticity and the
total energy of the magnetized star.   }
\end{abstract}

\end{titlepage}
\addtocounter{footnote}{-1}

\section{Introduction}

The Soft Gamma Repeaters are spectacular phenomena occurring in the
visible Universe. The giant flares detected so far show that the
peak luminosities are of order $10^{44} - 10^{46} erg/s$. One of the
most promising and  widely accepted explanations are the magnetars
\cite{DT}. Magnetars are believed to be neutron stars with ultra
strong magnetic field responsible for the observed giant flares. The
huge amount of energy released in the giant flares can be explained
by the existence of ultra strong magnetic fields with strength of
the order (or larger than) $10^{14} - 10^{15}$ Gauss \cite{H, K}.
The giant flares SGR 0526-66, SGR 1900+14 and SGR 1806-20 detected
so far reveal the existence of characteristic quasi periodic
oscillations  in the range  of tenths of Hz to kHz \cite{SW}. These
oscillations are believed to be seismic vibrations of the magnetars.
If the hypotheses is true this will provide  us with a tool to
investigate the stellar interior. That is why the quasi periodic
oscillations  were intensively studied in the past years \cite{L} --
\cite{C2} (and references therein).

The study of the stelar interior by the quasi periodic oscillations
require adequate models of the internal structure of the magnetars.
In general, our understanding of  magnetars as Soft Gamma Repeaters
is intimately related to the understanding their internal structure
and the construction of adequate models within General Relativity.
Clearly, the building of completely realistic magnetar models is a
formidable task. However, various simple relativistic models, more
or less realistic, could be built and these models provide us with
valuable physical insight into the internal structure of magnetars
\cite{BBGN} -- \cite{CFG}. The existing  simple magnetar models are
based on Einstein-Maxwell equations coupled to the perfect fluid
hydrodynamical equations. In modeling magnetar equilibrium
configurations two main approaches have been followed so far. The
first approach is to numerically solve the coupled systems of
equations \cite{BBGN},\cite{BG}, \cite{CPL}, \cite{KY}.  The second
approach is perturbative -- magnetar equilibrium configurations are
studied by using perturbative techniques, i.e.  the
Einstein-Maxwell-hydrodynamic equations are solved by linearizing
them about a known static and spherically symmetric background
solution of Einstein-hydrodynamic equations and then expanding the
perturbed equations in tensor harmonics \cite{KOK1},
\cite{KOK2},\cite{IS},\cite{CFGP} -- \cite{CFG}. Due to the linear
character of the perturbative equations one can consider in a
relatively simple manner more complicated magnetic field
configurations as the simultaneous presence of poloidal and toroidal
magnetic fields.

In the present paper we also address the problem of constructing
equilibrium configurations of neutron stars with ultra strong
magnetic fields within the framework of General Relativity. Contrary
to  the previous approaches mentioned above, our approach here is
fully analytical and  nonperturbative and based on exact solutions.
Exact solutions provide a route to better and  deeper understanding
of the inherent nonlinear character of gravity and its interaction
with matter. On the other hand, the exact solutions could serve as
tests for checking the computer codes which is important for the
advent of numerical relativity. More precisely, in this paper  we
find exact interior solutions to the coupled
Einstein-Maxwell-hydrodynamic equations describing static
(nonrotating) equilibrium configurations of strongly magnetized
neutron stars.  The interaction of the neutron star fluid with the
magnetic field is also taken into account to some extent.

\section{Setting of the problem and exact solutions}
Our starting point is the coupled Einstein-Maxwell-hydrodynamic
equations

\begin{eqnarray}
&&R_{\mu\nu}= 8\pi \left(T_{\mu\nu} -\frac{1}{2}Tg_{\mu\nu} \right)
+ 2 \left(F_{\mu\alpha}F_{\nu}^{\,\,\alpha}  - \frac{1}{4}F^2
g_{\mu\nu}\right) \\
&&\nabla_{\nu} F^{\mu \nu}=4\pi J^{\mu}, \\
&&\nabla_{[\mu} F_{\nu\alpha]}=0
\end{eqnarray}
where $T_{\mu\nu}$ and
$T_{\mu\nu}^{EM}=\frac{1}{4\pi}\left(F_{\mu\alpha}F_{\nu}^{\,\,\alpha}
- \frac{1}{4}F^2 g_{\mu\nu} \right)$ are the energy-momentum tensors
of the neutron matter and the electromagnetic field, respectively.
$J^{\mu}$ is the  current which sources the electromagnetic field.

Analytically solving of the coupled  Einstein-Maxwell-hydrodynamic
equations in the general case is a desperate task and therefore we
need some simplifying assumptions. We will assume that the
configurations (and the spacetime itself) are strictly static
(nonrotating) and axially symmetric. In mathematical terms our
assumptions mean that there exist one (hypersurface orthogonal)
timelike Killing vector $\xi$ and one spacelike axial Killing vector
$\eta$, commuting with $\xi$  and with closed periodic orbits
shrinking down to zero on the axis of symmetry. In adapted
coordinates the Killing vectors can be written in the usual form
$\xi=\partial/\partial t$ and $\xi=\partial/\partial \phi$ where $t$
is the time coordinate and $\phi$ is the azimuthal angle around the
axis of symmetry. Our geometrical assumptions impose restrictions on
the possible configurations of the electromagnetic field and the
energy momentum tensor of the neutron star matter. More precisely,
they require the absence of meridional convective currents and
electric field. The geometric assumptions require also the
4-velocity of the neutron matter to be aligned with the timelike
Killing vector $\xi$.

The invariance of the Maxwell 2-form $F$ under the axial Killing
field $\eta$  and the absence of meridional currents allow us to
introduce a magnetic potential $\Phi$ defined by $d\Phi=i_{\eta}F$.
The Maxwell 2-form then  is given by

\begin{eqnarray}
F= e^{-2u} \eta \wedge d\Phi,
\end{eqnarray}
where $e^{2u}=g(\eta,\eta)$. The magnetic field $B$ measured by a
comoving observer with 4-velocity $v^{\mu}$ is $B=i_{v}\star F$
where  $\star$ is the Hodge dual.

In the models studied so far the neutron matter has been described
by an isotropic perfect fluid with $T_{\mu\nu}=(\rho +
p)v_{\mu}v_{\nu} + pg_{\mu\nu}$ where $\rho$, $p$  and $v^{\mu}$ are
the energy density, the pressure and the 4-velocity of the fluid.
The description of the neutron star matter as an isotropic perfect
fluid is not completely satisfactory because it neglects the
interaction of the neutron matter with the ultra strong magnetic
field. This problem is highly nontrivial and extremely difficult to
be solved completely. From first principles it is clear that the
strong magnetic field yields anisotropy in the neutron star matter
and this should be taken into account in the energy-momentum tensor
of the matter. Indeed, since the neutron has anomalous magnetic
momentum the neutron matter will react to the strong magnetic field
by polarizing itself due to the coupling of the neutron spin to the
magnetic field. The things can get even more complicated if we take
into account the possible manifestation of some quantum effects like
the spin-spin interactions which can drive the system to some kind
of feromagnetic-like state. In this context we should also note that
the origin of the ultra strong magnetar magnetic fields is not
completely clear and   some sort of feromagnetic-like phase
transition could give contribution. Even more, the estimated
magnetic field on the magnetar surfaces mentioned above exceeds in
fact the QED critical magnetic field value $B_c\approx 10^{13} G$
which shows that the nonlinear Euler-Heisenberg electrodynamics
should be probably used instead of the linear Maxwell
electrodynamics. The above arguments show that the proper
description of the strong magnetic fields in the magnetars and the
properties of the neutron matter require subtle and extremely
complicated microscopic theory. The microscopic description of the
magnetars is far beyond the scope of this paper where we are
interested in the averaged macroscopic description which is
astrophysically relevant.

From a macroscopic point of view we can describe the interaction
(response) of the neutron matter with (to) the ultra strong magnetic
field by adding an anisotropic term to the energy-momentum tensor of
the isotropic perfect fluid. The anisotropy will manifest itself in
different pressures along the meridional planes of magnetic field
and in transverse direction. Indeed, according to the statistical
physics \cite{LL}, the pressure along the magnetic field is
$p=-\Omega$ while in transverse  direction $p^{tr}=-\Omega -
B{\mu}=p -B{\mu}$ where $\Omega$ is the grand canonical potential
and $\mu$ is the megnetization. In general the dependences
$\Omega(B)$ and $\mu(B)$ should be highly nonlinear and can be
determined only by the microscopic theory. In ultra strong magnetic
field, as we mentioned, most of the neutron spins should be oriented
in the direction of the magnetic field which means that $\mu>0$.
This shows that $p^{tr}<p$ in ultra strong magnetic field. As we
will see later the exact solutions predict the same behaviour for
the transverse pressure for realistic equations of state.

The only anisotropic term which  we can add and which is orthogonal
to the meridional planes and consistent with the geometrical
symmetries we imposed, is of the form $\sigma
e(\eta)_{\mu}e(\eta)_{\nu}$ where $\sigma$ is a scalar and
$e(\eta)^{\mu}$ is the unit vector along the axial Killing field
$\eta$. In other words we consider the following neutron star matter
energy-momentum tensor
\begin{eqnarray}\label{EMT1}
T_{\mu\nu}=(\rho + p)v_{\mu}v_{\nu} + pg_{\mu\nu} + \sigma
e(\eta)_{\mu}e(\eta)_{\nu}.
\end{eqnarray}
 The energy-momentum tensor (\ref{EMT1}) can be also
written in the form

\begin{eqnarray}
T=\rho v \otimes v + (p+\sigma)e(\eta) \otimes e(\eta) + p \left[g +
v \otimes v - e(\eta)\otimes e(\eta)\right]
\end{eqnarray}
which shows that $p$ is the fluid pressure in the meridional planes
where the magnetic field lays and $p^{tr}=p+\sigma$ is the pressure
in direction orthogonal to the meridional planes  and therefore
orthogonal to the magnetic field.

Armed with the energy-momentum tensor (\ref{EMT1}) we can write down
the dimensionally reduced  equations. Here we will perform the
dimensional reduction with respect to the spacelike axial Killing
vector $\eta$. For this purpose we need to introduce the
3-dimensional Lorentzian metric

\begin{eqnarray}
H= e^{2u}g - \eta \otimes \eta,
\end{eqnarray}
where $e^{2u}=g(\eta,\eta)$. The covariant derivative associated
with the metric $H$ will be  denoted by $D_i$. Then for the reduced
system of equations we obtain

\begin{eqnarray}
&&D_{i}D^{i}u= - 4\pi e^{-2u}(\rho - p) - e^{-2u} D_i\Phi D^{i}\Phi
- 4\pi \sigma e^{-2u}, \label{equ}\\ \nonumber
\\ &&{\cal R}(H)_{ij}= 8\pi (\rho + p) v_i
v_j + 8\pi (\rho-p)e^{-2u}H_{ij} + 2D_iu D_ju + 2e^{-2u} D_{i}\Phi
D_{j}\Phi, \label{eqR}
\\  \nonumber \\
&&D_{i}\left(e^{-2u}D^{i}\Phi\right)=4\pi
e^{-4u}J_{\phi},\label{eqJ}
\end{eqnarray}
along with the contracted Bianchi identity

\begin{eqnarray}
(\rho + p)D_{i} U + D_{i}p= -e^{-2u}J_{\phi} D_{i}\Phi + \sigma
D_{i} u. \label{bianchi}
\end{eqnarray}
Here ${\cal R}(H)_{ij}$ is the Ricci tensor with respect to the
3-metric $H_{ij}$, $e^{2U}=-g(\xi,\xi)$ and
$J_{\phi}=\eta^{\mu}J_{\mu}$.

Our main task now is to solve the system of coupled partial
differential equations (\ref{equ})--(\ref{bianchi}). Our strategy
for solving (\ref{equ})--(\ref{bianchi}) is to "add nonlinearly"
magnetic field to a known  static and  axisymmetric solution to
Einstein-hydrodynamic equations (i.e. without magnetic field)
described by the set $\{\rho^{0},p^{0},v_i^{0},u^{0}, H_{ij}^{0}\}$.
In order to do so we partially follow \cite{Y} where a method for
generating exact charged interior solutions was developed. We shall
assume that  $u$ and $\Phi$ depend on the space coordinates through
one function $\chi$, i.e. $u=u(\chi)$ and $\Phi(\chi)$. Substituting
into eq.(\ref{eqR}) we find

\begin{eqnarray}
{\cal R}(H)_{ij}= 8\pi (\rho + p) v_i v_j + 8\pi
(\rho-p)e^{-2u}H_{ij} + 2 \left[\left(\frac{du}{d\chi}\right)^2 +
e^{-2u}\left(\frac{d\Phi}{d\chi}\right)^2\right] D_{i}\chi
D_{j}\chi.
\end{eqnarray}

If we impose the relations

\begin{eqnarray}\label{new1}
H_{ij}=H^{0}_{ij},\;\; \chi=u^{0},\;\;  \rho=\rho^{0}
e^{2u(\chi)-2\chi}, \;\; p=p^{0} e^{2u(\chi)-2\chi}, \;\;
v_i=e^{\chi- u(\chi)} v_i^{0}
\end{eqnarray}
and

\begin{eqnarray}\label{new2}
\frac{d\Phi}{d\chi} = \pm
e^{u(\chi)}\sqrt{1-\left(\frac{du(\chi)}{d\chi}\right)^2}
\end{eqnarray}
we obtain that  eq.(\ref{eqR}) is automatically satisfied since
$\{\rho^{0},p^{0},v_i^{0},u^{0}, H_{ij}^{0}\}$ is a solution to the
static, axisymmetric Einstein-hydrodymanic equations by definition.
Then we can use eqs. (\ref{equ}) and (\ref{eqJ}) to find $\sigma$
and $J_{\phi}$:

\begin{eqnarray}
&&\sigma=
-(\rho^{0}-p^{0})e^{2u(\chi)-2\chi}\left(1-\frac{du(\chi)}{d\chi}\right)
-\frac{e^{2u(\chi)}}{4\pi}\left[\frac{d^2u(\chi)}{d\chi^2} +
e^{-2u(\chi)}\left(\frac{d\Phi(\chi)}{d\chi}\right)^2 \right]
D_i\chi D^i\chi, \label{sigma_new} \nonumber \\ \\
&& J_{\phi}= - \frac{d\Phi(\chi)}{d\chi} (\rho^{0}-p^{0})
e^{2u(\chi) - 2\chi} + \frac{e^{4u(\chi)}}{4\pi}
\frac{d}{d\chi}\left[e^{-2u(\chi)}\frac{d\Phi(\chi)}{d\chi}\right]D_i\chi
D^i\chi \label{J_new}.
\end{eqnarray}

It can be checked that eq. (\ref{bianchi}) is automatically
satisfied. Let us summarize the results in the following

\medskip
\noindent

{\bf Proposition.} {\it Let $\{\rho^{0},p^{0},v_i^{0},u^{0}=\chi,
H_{ij}^{0}\}$ be a solution to the Einstein-hydrodynamic equations
with isotropic perfect fluid and $u(\chi)$ is an arbitrary function
of $\chi$ with \\$\left(\frac{du(\chi)}{d\chi}\right)<1$. Then
$\{\rho, p, \sigma, v_i, H_{ij}=H_{ij}^{0}, u(\chi), \Phi(\chi),
J_{\phi}\}$ given by (\ref{new1}), (\ref{new2}), (\ref{sigma_new})
and (\ref{J_new}) form a solution to the
Einstein-Maxwell-hydrodynamic equations
(\ref{equ})--(\ref{bianchi}).}

\medskip
\noindent

This proposition allows to construct exact interior solutions with
arbitrary equation of state for the background solution. The only
exception is the case with stiff equation of state $\rho^{0}=p^{0}$
which is  very special and will not be considered here.

The 4-dimensional metric can be easily recovered form the data we
have. Namely, if

\begin{eqnarray}
ds^2_{0}= e^{2\chi}d\phi^2 + g^{0}_{ij}dx^idx^{j}
\end{eqnarray}
is the spacetime metric of the Einstein-hydrodynamic solution, then

\begin{eqnarray}
ds^2= e^{2u(\chi)}d\phi^2 + e^{-2u(\chi) +
2\chi}g^{0}_{ij}dx^idx^{j}
\end{eqnarray}
is the spacetime metric of the Einstein-Maxwell-hydrodynamic
solution, i.e of the magnetized solution.

From a physical point of view we have to impose some restrictions on
the functional dependence $u=u(\chi)$. More precisely, in order for
the new solution  to possess a well defined axis of symmetry the
function $u(\chi)$ should be of the form

\begin{eqnarray}\label{form_u}
u(\chi)=\chi + f(e^{2\chi}),
\end{eqnarray}
where $f(\chi)$ is a regular function with $f(0)=0$. In this way the
new solution will inherit the axis of symmetry from the background
solution used for its generation.

\section{Explicit exact solution}
Now we consider a physically interesting and realistic explicit
solution with $\sigma$ and $J_{\phi}$ vanishing on the star surface.
The solution is obtained by requiring $u(\chi)$ and $\Phi(\chi)$ to
satisfy the equations of the affinely parameterized geodesics of the
2-dimensional metric $dl^2=du^2 + e^{-2u}d\Phi^2$, i.e. the
equations

\begin{eqnarray}
&&\frac{d^2u(\chi)}{d\chi^2} +
e^{-2u(\chi)}\left(\frac{d\Phi(\chi)}{d\chi}\right)^2 =0, \\
&&\frac{d}{d\chi}\left[e^{-2u(\chi)}\frac{d\Phi(\chi)}{d\chi}\right]=0.
\end{eqnarray}

This requirement considerably simplifies equations (\ref{sigma_new})
and (\ref{J_new}). The solution of the above equations is

\begin{eqnarray}
&&e^{2u(\chi)}= \frac{e^{2\chi}}{\left(1 + b^2 e^{2\chi}\right)^2},\\
&&\Phi(\chi)= b \frac{e^{2\chi}}{1 + b^2 e^{2\chi}},
\end{eqnarray}
where $b$ is an arbitrary parameter\footnote{The other parameter has
been appropriately chosen in order to have a well defined axis.  }.
One can see that $u(\chi)$ is of the form (\ref{form_u}) and
therefore the solution has a well defined axis of symmetry. The
physical meaning of the parameter $b$ can be uncovered as follows.
For the strength of the magnetic field we have

\begin{eqnarray}
{\vec B}^2= \frac{1}{2} F^2 = e^{-2\chi}
\left(\frac{d\Phi(\chi)}{d\chi}\right)^2 g^{0\,ij}\partial_{i}\chi
\partial_{j}\chi=  \frac{4b^2}{\left(1 +
b^2e^{2\chi}\right)^4}  g^{0\,ij}
\partial_{i}e^{\chi}\partial_{j}e^{\chi}.
\end{eqnarray}

Taking into account the space is locally Euclidian in small
neighborhood of the axis and the fact that $e^{2\chi}|_{axis}=0$, it
is not difficult to find the strength $B_{0}$ of the magnetic field
on the axis
\begin{eqnarray}
B_{0}^2= 4b^2.
\end{eqnarray}
In fact $B_{0}$ is also the strength  of the magnetic field on the
north or south pole of the star surface. So the parameter $b$ can be
interpreted as being  one half of the north pole magnetic field
strength, i.e. $b=\frac{1}{2}B_{0}$.

In order to be more specific we will consider a spherically
symmetric background solution. Also we will present the background
solution in the widely used Schwarzschild coordinates $r$ and
$\theta$ with

\begin{eqnarray}
g^{0}_{\theta\theta}= r^2, \;\;
g^{0}_{\phi\phi}=e^{2\chi}=r^2\sin^2\theta.
\end{eqnarray}

Then our magnetized solution is as follows

\begin{eqnarray}
&&ds^2 =  \Lambda^{2} \left(g^{0}_{tt}dt^2 + g^{0}_{rr}dr^2 +
r^{2}d\theta^2\right) + \Lambda^{-2} r^2\sin^2\theta d\phi^2, \\\nonumber  \\
&&\rho= \Lambda^{-2}\rho^{0}, \;\;  p= \Lambda^{-2}p^{0}, \\ \nonumber \\
&&\Phi= \frac{1}{2}\Lambda^{-1}B_{0}r^2\sin^2\theta, \\ \nonumber  \\
&&\sigma = -\frac{1}{2}B^2_{0} \Lambda^{-3}(\rho^{0}-p^{0})
r^2\sin^2\theta \label{sigma_ss},\\ \nonumber \\
&&J_{\phi}= - B_{0}\Lambda^{-4} (\rho^{0}-p^{0})r^2\sin^2\theta,
\label{J_ss}
\end{eqnarray}
where
\begin{eqnarray}
\Lambda= 1 + \frac{1}{4}B^2_{0} r^2\sin^2\theta.
\end{eqnarray}
The nonzero components of the magnetic field are

\begin{eqnarray}
&&B_{r}= -B_{0}\Lambda^{-1}\sqrt{g^{0}_{rr}} \cos\theta,\\ \nonumber \\
&&B_{\theta} =
B_{0}\Lambda^{-1}\frac{r\sin\theta}{\sqrt{g^{0}_{rr}}}.
\end{eqnarray}

We see that when the background solution has a well defined boundary
at $r=R$ corresponding to the star surface where $p^{0}(R)=0$, the
same is true for the magnetized solution, i.e.  $p(R)=0$  since
$p=\Lambda^2 p^{0}$. Moreover, if $\rho^{0}$ also vanishes on the
star surface the same holds for $\sigma$ and $J_{\phi}$ according to
(\ref{sigma_ss}) and (\ref{J_ss}). As we should expect the
anisotropy pressure $\sigma$  is yielded by the magnetic field and
vanishes for zero magnetic field. Also, as we discussed in Section
2, the transverse pressure $p^{tr}=p + \sigma$ should be smaller
than $p$ in strong magnetic field. Indeed we see that for realistic
equations of state for the background solutions, i.e. for
$\rho^{0}\ge p^{0}$ we have $\sigma\le 0$.

In order to describe the way in which the magnetic field deforms the
star we consider the space metric on the star surface, namely

\begin{eqnarray}
dl^2_s= R^2 \left(\Lambda_{s}^2 d\theta^2 + \Lambda_{s}^{-2}
\sin^2\theta d\phi^2\right),
\end{eqnarray}
where $\Lambda_s=1 + \frac{1}{4}B^2_{0}R^2\sin^2\theta$. The
circumference about the equator ($\theta=\pi/2$) is

\begin{eqnarray}
L_{e}=\int^{2\pi}_{0} \Lambda_{s}^{-1} R d\phi = \frac{2\pi R}{1 +
\frac{1}{4}B^2_{0}R^2},
\end{eqnarray}
while for the polar circumference ($\phi=const$) we have

\begin{eqnarray}
L_{p}= 2\int^{\pi}_{0} \Lambda_{s} R d\theta= 2\pi R \left (1
+\frac{1}{8}B^2_{0}R^2 \right).
\end{eqnarray}

The surface elipticity $\varepsilon_{surf}$ is given by
\begin{eqnarray}
\varepsilon_{surf}= \frac{L_e - L_p}{L_p}
\end{eqnarray}
and $\varepsilon_{surf}<0$ for $B_{0}\ne 0$. Therefore, for the
solution under consideration  the magnetic field elongates the star
along the magnetic field -- the star is prolate in shape\footnote{It
is worth noting that the negative surface elipticity is a
characteristic of the specific solution we consider. In principle
the more general solutions which can be generated via our
proposition  may  give positive elipticity. In those cases,
however,the current $J_{\phi}$ and anisotropy pressure $\sigma$ do
not vanish on the star surface}. For small $B^2_{0}R^2$ we have
$\varepsilon_{surf}\approx -\frac{3}{8}B^2_{0}R^2$. Here we should
note that the numerical and perturbative models with pure poloidal
magnetic field predict positive surface elipticity. The reason for
that discrepancy is the fact that  the perturbative and numeric
models consider the neutron star matter as pure isotropic perfect
fluid without taking into account the anisotropy caused by the
interaction with magnetic field.

The next physical quantity we shall consider is the total energy $M$
concentrated in the star

\begin{eqnarray}
M= - \frac{1}{4\pi} \int_{Star} R^{t}_{t}\sqrt{-g}d^3x =
\int_{Star}\left(\rho + 3p + \sigma + \frac{1}{4\pi}{\vec B}^2
\right)\sqrt{-g}d^3x.
\end{eqnarray}

Taking into account (\ref{new1}) and $(\ref{sigma_new})$ we find

\begin{eqnarray}\label{Mass}
M= M_{0} + \frac{1}{2} B^2_{0} \int^{R}_{r=0}\int^{\pi}_{\theta=0}
\left[\Lambda^{-2}\!\left(\frac{\sin^2\theta}{g^{0}_{rr}} +
\cos^2\theta\!\right) \right. \\ \nonumber \\\left. -
2\pi\Lambda^{-1}(\rho_0 -p_{0})r^2\sin^2\theta
\right]\!\sqrt{|g^{0}_{tt}|g^{0}_{rr}}r^2\sin\theta drd\theta
\nonumber
\end{eqnarray}
where
\begin{eqnarray}
M_{0}=\int_{Star}\left(\rho_{0} + 3p_{0}\right)\sqrt{-g^{0}}d^3x
\end{eqnarray}
is the total energy of the background solution. The explicit form of
$M$ depends of the background solution but we can give a good
approximation by using the interior Schwarzschild solution as  a
representative example of a background solution.  The interior
Schwarzschild solution is characterized by a constant energy density
$\rho_{0}=3 M_{0}/4\pi R^3$ and the metric and pressure are given by

\begin{eqnarray}
&&ds^2_{0}= -\!\left[{3\over 2}\left( 1- {2M_{0}\over R}
\right)^{1/2} \!\!- {1\over 2}\left(1 - {2M_{0}\over
R^3}r^2\right)^{1/2} \right]^2\! dt^2 + \frac{dr^2}{1 -
\frac{2M_{0}}{R^3}r^2} +
r^2 \left(d\theta^2 + \sin^2\theta d\phi^2 \right) , \nonumber \\ \\
&&p_{0} = {3M_{0}\over 4\pi R^3} \left[\left(1-
\frac{2M_{0}}{R^3}r^2\right)^{1/2} - \left(1-  \frac{2M_{0}}{R}
\right)^{1/2}\over 3\left(1-  \frac{2M_{0}}{R} \right)^{1/2} -
\left(1- \frac{2M_{0}}{R^3}r^2\right)^{1/2}\right]. \nonumber
\end{eqnarray}
The mass-radius ratio satisfies the inequality $2M_{0}/R<8/9$.
Substituting the interior Schwarzschild solution into (\ref{Mass})
and performing calculations up to terms in the order $B^2_{0}R^2$,
we find

\begin{eqnarray}
M=M_{0} + \frac{1}{3} B^2_{0} R^3\left(1-
\frac{1}{5}B_{0}^2R^2\right) \left(1- \frac{2M_{0}}{R}\right) +
{\cal O}\left((B_{0}R)^4\right).
\end{eqnarray}

\section{Conclusion}

In this paper we presented a simple method for generating exact
interior solutions to the static and axisymmetric
Einstein-Maxwell-hydrodynamic equations with anisotropic perfect
fluid. On this base we can build simple nonperturbative analytical
relativistic models of the magnetars. To the best of our knowledge
these are the first nonperturbative analytical relativistic models
of the magnetars with arbitrary equation of state. As an
illustration we gave an explicit realistic exact interior solution
for the magnetars and on its base we  calculated the suraface
elipticity of the star and its energy.

The present work could be extended in several directions. It would
interesting and important more general configurations of the
magnetic field, i.e. a mixture of poloidal amd toroidal magnetic
fields   to be investigated. The next interesting extension is to
add rotation to the star. The mentioned possible extensions are very
challenging due to the highly nonlinear character of the Einstein
equations. We hope, however, that some progress could be made.

\vspace{1.5ex}
\begin{flushleft}
\large\bf Acknowledgments
\end{flushleft}

The author would like to thank K. Kokkotas and D. Doneva for reading
the manuscript and the discussions. The author would like to thank
the Alexander von Humboldt Foundation for the support, and the
Institut f\"ur Theoretische Astrophysik T\"ubingen for its kind
hospitality. He also acknowledges partial financial support from the
Bulgarian National Science Fund under Grant DO 02-257 and by Sofia
University Research Fund under Grant 88/2011.

\end{document}